\documentclass[preprint,preprintnumbers,amsmath,amssymb]{revtex4}
\usepackage{dcolumn}
\usepackage{bm}
\usepackage{graphicx}

\begin{document}

\title{Strong cosmic censorship for the massless charged scalar field in the Reissner-Nordstrom-de Sitter spacetime}
\author{Yuyu Mo$^{1,2}$}
\thanks{yymo@mail.bnu.edu.cn}
\author{Yu Tian$^{3}$}
\thanks{ytian@ucas.ac.cn}
 \author{Bin Wang$^{1,4}$}
 \thanks{wangb@yzu.edu.cn}
\author{Hongbao Zhang$^{2,5}$}
\thanks{hzhang@vub.ac.be}
 \author{Zhen Zhong$^{2}$}
 \thanks{zhenzhong@mail.bnu.edu.cn}

\affiliation{
 $^1$ Center for Gravitation and Cosmology, College of Physical Science and Technology,
Yangzhou University, Yangzhou 225009, China\\
$^2$ Department of Physics, Beijing Normal University, Beijing 100875, China\\
$^3$ School of Physics, University of Chinese Academy of Sciences, Beijing 100049, China\\
$^4$ Department of Physics and Astronomy, Shanghai Jiao Tong University,
Shanghai 200240, China\\
 $^5$ Theoretische Natuurkunde, Vrije Universiteit Brussel,
 and The International Solvay Institutes, Pleinlaan 2, B-1050 Brussels, Belgium
}

\begin{abstract}
It has recently been shown that the strong cosmic censorship conjecture can be violated by the massless neutral scalar field in the nearly extremal Reissner-Nordstrom-de Sitter black hole. However, the formation of such a black hole by gravitational collapse necessitates the presence of the charged sector on top of the Einstein-Maxwell system. Thus we numerically calculate the quasi-normal modes for a massless charged scalar field in the Reissner-Nordstrom-de Sitter spacetime by generalizing the characteristic formulation to the charged case. As a result, the strong cosmic censorship turns out to be recovered by our massless charged scalar field except in the highly extremal limit $Q\rightarrow Q_m$, where the violation still occurs when the scalar field is appropriately charged.

\end{abstract}

\maketitle

\section{Introduction and motivation}
As is well known, a variety of versions of singularity theorems tell us that spacetime singularity can be formed generically by the gravitational collapse of suitable matter distribution\cite{Penrose1,HP}. Such a formation of singularity indicates that general relativity breaks down near the singularity, and will be replaced by the so-called complete quantum theory of gravity. In particular, if the formed singularity is timelike, general relativity will lose its predictive power because there exists the region of spacetime in consideration, whose past domain of influence  will hit the singularity. With this in mind, Penrose proposed his strong cosmic censorship hypothesis (SCC) to maintain the predictability of classical general relativity. SCC, in essence, states that the gravitational collapse of generic initial distribution for suitable matter leads only to a spacelike or lightlike singularity.

The timelike singularity in the eternal Kerr and Reissner-Nordstrom black holes appears to violate SCC. However, this is not the case. Because the remnant fields are generically present along with the real life black hole formed from the gravitational collapse. These remnant fields, which are demonstrated to have an inverse power-law decay behavior outside of the black hole, will be amplified when propagated along the Cauchy horizon due to the exponential blue-shift effect over there. As a result, the Cauchy horizon becomes singular such that one cannot extend across the would-be Cauchy horizon to the spacetime region with the timelike singularity.

But the above argument does not apply to the black holes in de Sitter spacetime because the remnant fields instead have an exponential decay behavior outside of the black hole\cite{Chambers,BCKL,Dafermos,CGNS,HV,CF,Hintz}. Accordingly, the extendibility of the Cauchy horizon depends delicately on the competition between the exponential decay behavior outside of the black hole and the exponential blue-shift amplification along the Cauchy horizon. As shown in \cite{DERS}, the blue-shift effect wins for the remnant fields around the Kerr-de Sitter black hole, so SCC is respected. On the other hand, when one considers the remnant massless neutral scalar field around Reissner-Nordstrom-de Sitter (RNdS) black hole, the exponential decay effect wins in some regime of the parameter space in consideration such that SCC is violated\cite{CCDHJ}. Such a violation even becomes more severe for the coupled electromagnetic and gravitational perturbations\cite{DRS1}. However, taking into account the unavoidable presence of charged remnant fields in the dynamical formation of the RNdS black hole, Hod finds that SCC is restored at least by the scalar field with a sufficiently large charge and mass\cite{Hod}. The purpose of this paper is to see what happens to SCC if one charges the massless scalar field considered in \cite{CCDHJ}. In this case, Hod's analytic analysis does not work, so we are required to numerically calculate the low-lying quasi-normal modes (QNMs) for the massless charged scalar field in RNdS black hole background. Our result shows that SCC is recovered by our massless charged scalar field except in the highly extremal limit, where the violation can still occurs when the charge of our scalar field is tuned to some appropriate regime.

The structure of this paper is organized as follows. In the subsequent section, we develop the relationship between the QNMs and SCC for the charged scalar field in RNdS background. In Section \ref{numerics}, after introducing our numerical scheme for the time evolution of the charged scalar field by the double null coordinates, we present the relevant numerical results about the low-lying QNMs for the massless charged scalar field and the implications to SCC. We conclude our paper in the last section with some discussions.
\section{Quasi-normal modes and strong cosmic censorship}
Let us start with the four dimensional RNdS black hole
\begin{equation}
ds^2=-f(r)dt^2+\frac{dr^2}{f(r)}+r^2(d\theta^2+\sin^2\theta d\phi^2), \quad A_a=-\frac{Q}{r}(dt)_a,
\end{equation}
where the the blackening factor
\begin{equation}
f(r)=1-\frac{2M}{r}+\frac{Q^2}{r^2}-\frac{\Lambda r^2}{3}
\end{equation}
with $M$ and $Q$ the mass and charge of the black hole, and $\Lambda$ the positive cosmological constant. If the cosmological, event, and Cauchy horizons are designated as $r_c$, $r_+$, and $r_-$ individually, then the blackening factor can also be written as
\begin{equation}
f(r)=\frac{\Lambda}{3r^2}(r_c-r)(r-r_+)(r-r_-)(r-r_o)
\end{equation}
with $r_o=-(r_c+r_++r_-)$. In addition, the surface gravity at each horizon $r_h$ is given by $\kappa_h=|\frac{1}{2}f'(r_h)|$. Whence we have
\begin{eqnarray}
\kappa_c&=&\frac{\Lambda}{6r_c^2}(r_c-r_+)(r_c-r_-)(r_c-r_o),\nonumber\\
\kappa_+&=&\frac{\Lambda}{6r_+^2}(r_c-r_+)(r_+-r_-)(r_+-r_o),\nonumber\\
\kappa_-&=&\frac{\Lambda}{6r_-^2}(r_c-r_-)(r_+-r_-)(r_--r_o),\nonumber\\
\kappa_o&=&\frac{\Lambda}{6r_o^2}(r_c-r_o)(r_+-r_o)(r_--r_o).
\end{eqnarray}

Now suppose that the behavior of the charged scalar field in such a curved spacetime is governed by the following Klein-Gordon equation
\begin{equation}
[(\nabla_a-iqA_a)(\nabla^a-iqA^a)-m^2]\Psi=0,
\end{equation}
which can be written explicitly as
\begin{equation}
-\frac{\partial_t^2\Psi}{f}+\frac{1}{r^2}\partial_r(r^2f\partial_r\Psi)+\frac{1}{r^2}[
\frac{1}{\sin\theta}\partial_\theta(\sin\theta\partial_\theta\Psi)+\frac{1}{\sin^2\theta}\partial_\phi^2\Psi]
-\frac{2iqQ}{rf}\partial_t\Psi+\frac{(qQ)^2}{r^2f}\Psi-m^2\Psi=0
\end{equation}
with $m$ and $q$ the mass and charge of the scalar field. Associated with an arbitrary solution to this equation of motion, not only is there a gauge transformation $(A_a, \Psi)\rightarrow (A_a+\nabla_a\lambda, e^{iq\lambda}\Psi)$, but also a conserved current given by
\begin{equation}
j^a=i[\bar{\Psi}(\nabla^a-iqA^a)\Psi-\Psi(\nabla^a+iqA^a)\bar{\Psi}].
\end{equation}
Due to the symmetry of the background and the linearity of the dynamics, it is sufficient for us to consider the scalar field as
\begin{equation}
\Psi=\frac{\psi(r)}{r}Y_{lm}(\theta,\phi)e^{-i\omega t}.
\end{equation}
Plugging it into the above equation of motion, we wind up with the effective equation
\begin{equation}
\frac{d^2\psi}{dr_*^2}+\{[\omega-\Phi(r)]^2-V(r)\}\psi=0
\end{equation}
for the radial function, where the tortoise coordinate $r_*$ is defined as $dr_*=\frac{dr}{f}$ with the electric potential $\Phi(r)=\frac{qQ}{r}$ and the effective potential $V(r)=\frac{f[rf'+l(l+1)+m^2r^2]}{r^2}$. It obviously follows that the radial function behaves as
\begin{equation}
\psi\sim e^{\pm i[\omega-\Phi(r_h)]r_*}
\end{equation}
near any one of the horizons $r_h$.
Now let us consider the regime between the event and cosmological horizons, where $r_*$ can be integrated out as
\begin{equation}
r_*=-\frac{1}{2\kappa_c}\ln(1-\frac{r}{r_c})+\frac{1}{2\kappa_+}\ln(\frac{r}{r_+}-1)-\frac{1}{2\kappa_-}\ln(\frac{r}{r_-}-1)+\frac{1}{2\kappa_o}\ln(1-\frac{r}{r_o}).
\end{equation}
If we impose the following boundary conditions
\begin{equation}
\psi\sim e^{-i[\omega-\Phi(r_+)]r_*} \quad r\rightarrow r_+, \quad \psi\sim e^{i[\omega-\Phi(r_c)]r_*}\quad  r\rightarrow r_c,
\end{equation}
namely, the ingoing boundary condition near the event horizon and the outgoing boundary condition near the cosmological horizon, then the equation of motion will give rise to a set of discrete frequencies, which is the so-called QNMs. It is noteworthy that the spectrum of QNMs for the scalar field with the charge $q$ is related to that with the charge $-q$ by the minus complex conjugation. So when the scalar is neutral, the spectrum of QNMs is symmetric with respect to the imaginary axis on the $\omega$-plane. As we see in the next section, this symmetry will be broken at the presence of the charge.

On the other hand, with the above ingoing boundary condition near the event horizon, the solution inside of the black hole, however, generically has both the outgoing and ingoing modes near the Cauchy horizon. By performing the gauge transformation with $d\lambda=\frac{Q}{r}dr_*$, and the coordinate transformation to the outgoing coordinates with $u$ defined as $u=t-r_*$, which allows us to analytically continue our metric and electric potential across the Cauchy horizon, the outgoing and ingoing modes can be expressed as
\begin{equation}
\psi_o \sim e^{-i\omega u}, \quad \psi_i \sim e^{-i\omega u}(r-r_-)^\frac{i[\omega-\Phi(r_-)]}{\kappa_-}, \quad r\rightarrow r_-
\end{equation}
near the Cauchy horizon, respectively. Obviously, the potential non-smoothness near the Cauchy horizon comes from the ingoing mode $\psi_i$. As detailed in \cite{DERS}, one can extend this mode across the Cauchy horizon such that SCC is violated if and only if it has a locally square integrable derivative, belonging to the Sobolev space $H^1_{loc}$, which requires
\begin{equation}
\beta\equiv-\frac{\text{Im}(\omega)}{\kappa_-}> \frac{1}{2}.
\end{equation}
Put it another way, if one can find a quasi-normal mode with $\beta <\frac{1}{2}$, then SCC is preserved. So for this purpose, we are only required to focus on the lowest-lying quasi-normal mode.
\begin{figure}
  \centering
  \includegraphics[width=6.0cm]{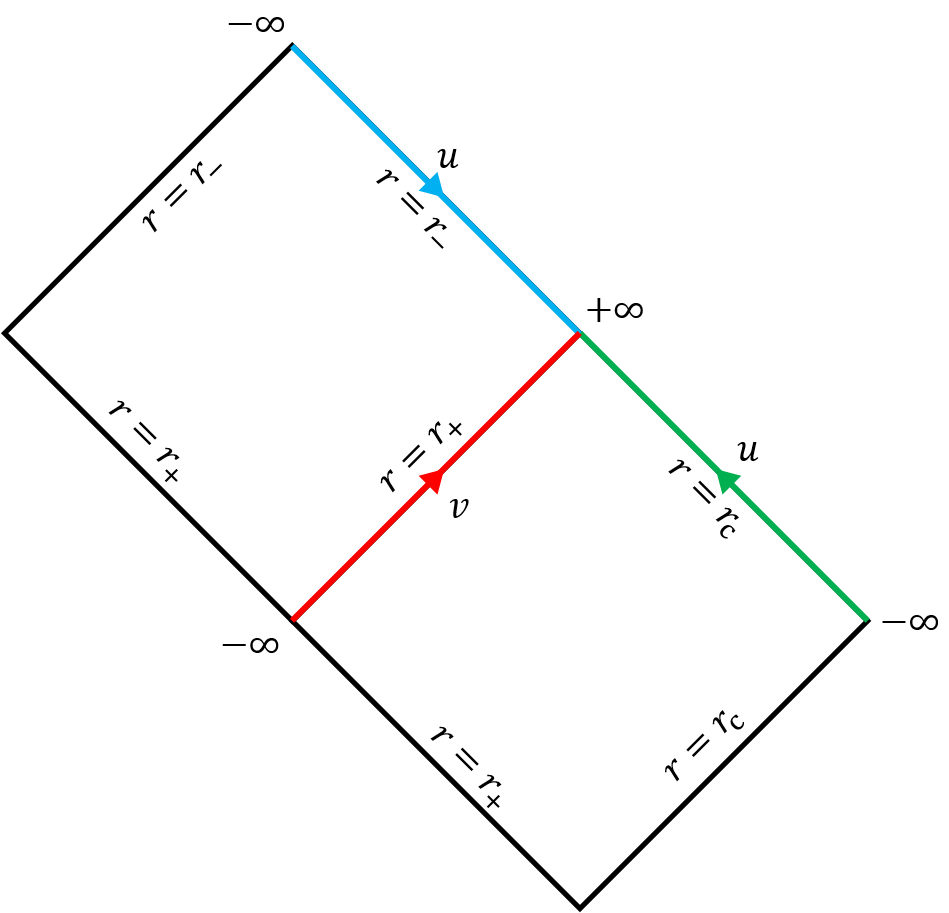}\\
  \caption{Penrose diagram for the RNdS black hole with $u$ and $v$ defined as $u=t-r_*$ and $v=t+r_*$. The red line, blue line, and green line denote the black hole event horizon, Cauchy horizon, and cosmological horizon individually.}\label{PD}
\end{figure}

\section{Numerical scheme and relevant results}\label{numerics}
\subsection{Numerical scheme}
Regarding the QNMs of the charged scalar field in the RNdS black hole, early works include \cite{ZZPWA} and \cite{KZ}, where the main issue they focused on is the superradiance instability and the numerical scheme they used is the general initial value integration. Below we shall also extract the QNMs by the time domain analysis, but instead we shall evolve our initial data by the characteristic formulation, which was proposed originally in \cite{GPP}. This characteristic formulation has been adopted in various case studies\cite{BCLP,WMA,MGAS,WLM,LMRT}. We shall generalize this formulation to our charged scalar field. To proceed, as demonstrated in FIG.\ref{PD}, we first make the coordinate transformation to the double null coordinates $(u,v)$ with $u$ defined before and $v$ defined as $v=t+r_*$. Accordingly the metric reads
\begin{equation}
ds^2=-fdudv+r^2(d\theta^2+\sin^2\theta d\phi^2).
\end{equation}
In addition, we would also like to make the gauge transformation $d\lambda=\frac{Q(2r-r_c-r_+)}{r(r_c-r_+)}dr_*$
such that the electric potential reads
\begin{equation}
A_a=-\frac{Q(r-r_+)}{r(r_c-r_+)}(du)_a+\frac{Q(r-r_c)}{r(r_c-r_+)}(dv)_a.
\end{equation}
If we expand the scalar field as
\begin{equation}
\Psi=\frac{\psi(u,v)}{r}Y_{lm}(\theta,\phi),
\end{equation}
then the resultant Klein-Gordon equation can be expressed as
\begin{equation}
0=-4\partial_u\partial_v\psi-4i\Phi(r)(\frac{r_c-r}{r_c-r_+}\partial_u\psi+\frac{r-r_+}{r_c-r_+}\partial_v\psi)-U(r)\psi,
\end{equation}
where $U(r)=\frac{4\Phi^2(r)(r-r_c)(r-r_+)}{(r_c-r_+)^2}+\frac{f}{r^2}[(l(l+1)+f'r+m^2r^2+i\frac{\Phi(r)r(r_c+r_+)}{r_c-r_+}]$.
Whence it is not hard to see the QNMs behave as $e^{-i\omega v}$ near $r_+$ and $e^{-i\omega u}$ near $r_c$. This is actually the reason why we have made the above gauge transformation.
\begin{figure}
  \centering
  \includegraphics[width=6.0cm]{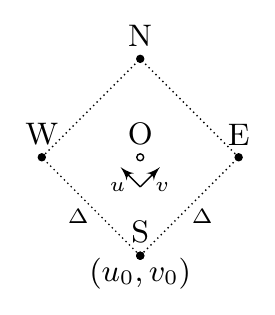}\\
  \caption{Finite difference scheme to obtain the data at $N$ from the data at $E$, $W$, and $S$.}\label{FD}
\end{figure}

\begin{figure}
  \centering
  \includegraphics[width=6.0cm]{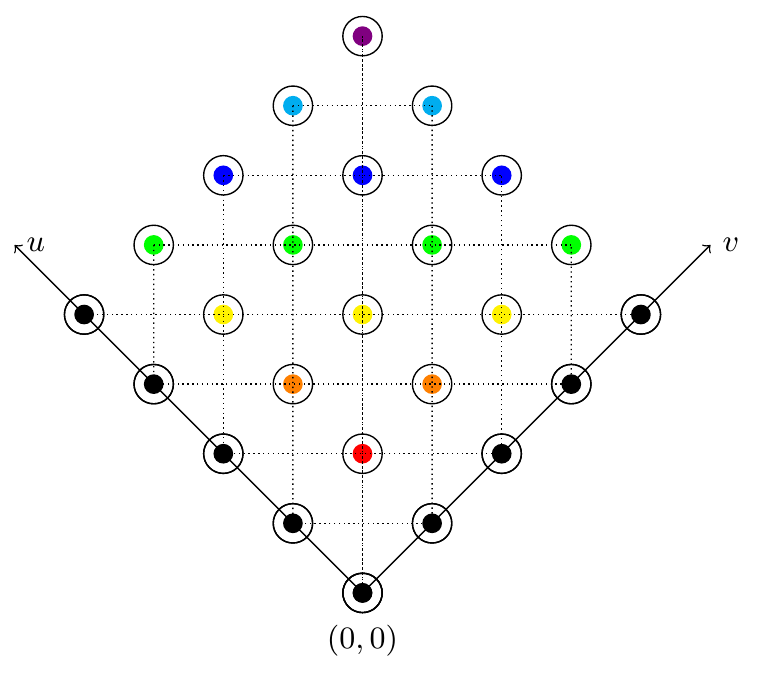}\\
  \caption{Diamond parallel evolution scheme, where horizontal and vertical lines correspond to the time $t$ slice and equal $r$ line, respectively.}\label{DE}
\end{figure}

To numerically solve the above partial differential equation, we would like to approximate it at the point $O(u_0+\frac{\Delta}{2},v_0+\frac{\Delta}{2})$ to the order of $O(\Delta^2)$ in the following way
\begin{eqnarray}
\partial_u\partial_v\psi|_O &\rightarrow& \frac{\psi(N)-\psi(E)-\psi(W)+\psi(S)}{\Delta^2},\\
\partial_u\psi|_O&\rightarrow&\frac{\psi(N)+\psi(W)-\psi(E)-\psi(S)}{2\Delta},\\
\partial_v\psi|_O&\rightarrow&\frac{\psi(N)+\psi(E)-\psi(W)-\psi(S)}{2\Delta},\\
\psi(O)&\rightarrow&\frac{\psi(E)+\psi(W)}{2},
\end{eqnarray}
where as depicted in FIG.\ref{FD}, $N$, $E$, $W$, and $S$ correspond individually to the points $(u_0+\Delta,v_0+\Delta)$, $(u_0,v_0+\Delta)$, $(u_0+\Delta,v_0)$, and $(u_0,v_0)$. With this approximation, the equation of motion gives rise to
\begin{eqnarray}
\psi(N)&=&(1+i\frac{\Phi(r)\Delta}{2})^{-1}[-(1-i\frac{\Phi(r)\Delta}{2})\psi(S)\\
&&-i\frac{(2r-r_c-r_+)\Phi(r)\Delta}{2(r_c-r_+)}(\psi(E)-\psi(W))+(1-\frac{U(r)\Delta^2}{8})(\psi(E)+\psi(W))],
\end{eqnarray}
where $r$ is evaluated at the point $O$, thus can be solved by $r_*(r)=\frac{1}{2}(v_0-u_0)$. As illustrated in FIG.\ref{DE}, to reduce the computing time, we adopt a parallel evolution along the time $t$ slice within the diamond in consideration. In addition, in our numerical evolution, the initial value for our scalar field is set as follows
\begin{eqnarray}
\psi(0,v)&=&0,\nonumber\\
\psi(u,0)&=&\frac{1}{\sqrt{2\pi}\sigma}e^{-\frac{(u-u_c)^2}{2\sigma^2}}
\end{eqnarray}
with $\sigma$ and $u_c$ as the width and center of the Gaussian wave packet.
Then we shall extract the spectrum of low-lying QNMs from the $N$ equally elapsed late time data $\hat{\psi}(t_p)=\psi(t_0+p\Delta,r_*=0)$ by Prony method\cite{BCGS}. The convergence of our numerics is examined by decreasing the evolution step length $\Delta$. We have also tested our numerics by reproducing the relevant results reported in the previous literatures such as \cite{CCDHJ,ZZPWA,KZ}. Below we shall focus only on the massless scalar field, although the afore-prescribed numerical scheme can be applied equally to the massive case. In addition, we shall work with the units in which $M=1$.
\begin{figure}
  \centering
  \includegraphics[width=12.0cm]{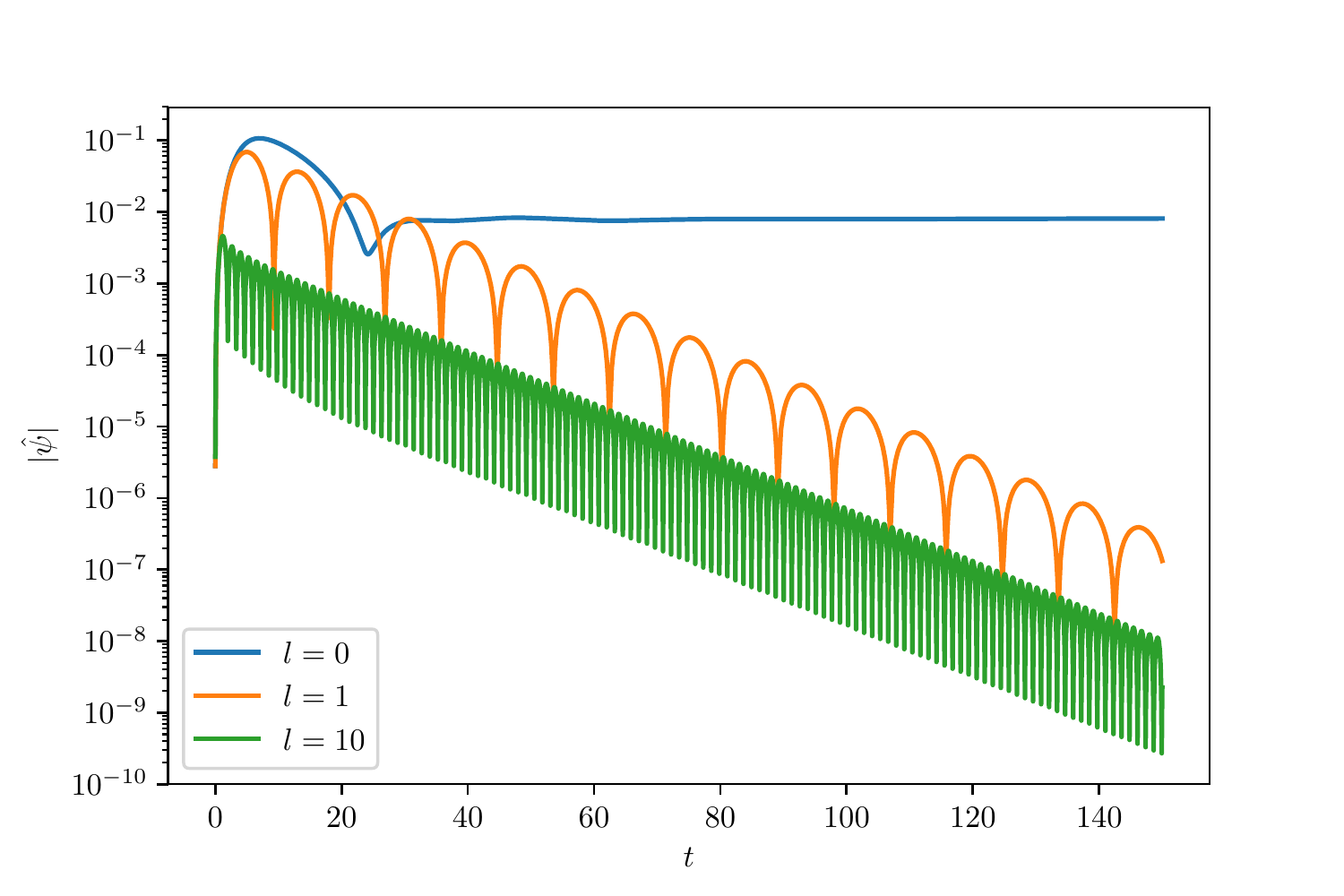}\\
  \caption{The temporal evolution of $|\hat{\psi}(t)|$ of $q=0.1$ for $\Lambda=0.02$ and $Q/Q_m=0.9910$.}\label{temporal}
\end{figure}

As a demonstration, we would like to conclude this subsection by depicting the temporal evolution of $\hat{\psi}(t)$ of $q=0.1$ in FIG.\ref{temporal} and list the corresponding spectrum of low-lying QNMs in TABLE I for $\Lambda=0.02$ and $Q/Q_m=0.9910$, where $Q_m$ corresponds to the charge of black hole with $r_+=r_-$ and $n$ denotes the overtone number with $n=1$ representing the fundamental mode. Among others, we see there is a slowly growing unstable mode for $l=0$ with its real part $\frac{\text{Re}(\omega)}{\kappa_-}\in(\frac{\Phi(r_c)}{\kappa_-}=0.051966,\frac{\Phi(r_+)}{\kappa_-}=0.503183)$. Actually, as shown before in \cite{ZZPWA,KZ}, this kind of unstable mode appears only for $l=0$ and is always superradiant.

\begin{table}
\caption{The low-lying QNMs $\frac{\omega}{\kappa_-}$ of $q=0.1$ for $\Lambda=0.02$ and $Q/Q_m=0.9910$}\label{table}
\begin{tabular}{@{}p{1cm}p{4cm}p{4cm}p{4cm}@{}}\toprule
$n$ & $l=0$ & $l=1$ & $l=10$ \\ \colrule
$1$ & $0.057773+0.002227i$ & $0.032203-0.475118i$ & $-14.080-0.491i$ \\
$2$ & $1.008291-0.520028i$ & $2.348530-0.499977i$ & $14.653-0.492i$ \\
$3$ & $-0.452819-0.559594i$ & $-1.769203-0.500762i$ & $-14.059-1.474i$\\
$4$ & $0.598660-0.834292i$ & $-0.033667-1.435223i$ & $14.633-1.476i$\\\botrule
\end{tabular}
\end{table}

\subsection{Relevant results}
Due to the limited computational resources, we have no intention to chart the corresponding spectrum of low-lying QNMs for the whole parameter space. Actually, as demonstrated in \cite{CCDHJ}, only in the near extremal RNdS black hole is SCC violated by the massless neutral scalar field, where the dominant criminal modes can be the de Sitter mode, photon sphere mode, or near extremal mode, depending on the specific parameter value in the moduli space. With this in mind, we simply investigate some representative points in the moduli space, which also suffices for our purpose. In particular, the following representative points are so chosen as to facilitate the comparison, if necessary, with the result presented in \cite{CCDHJ}.

\begin{figure}
  \centering
  \includegraphics[width=12.0cm]{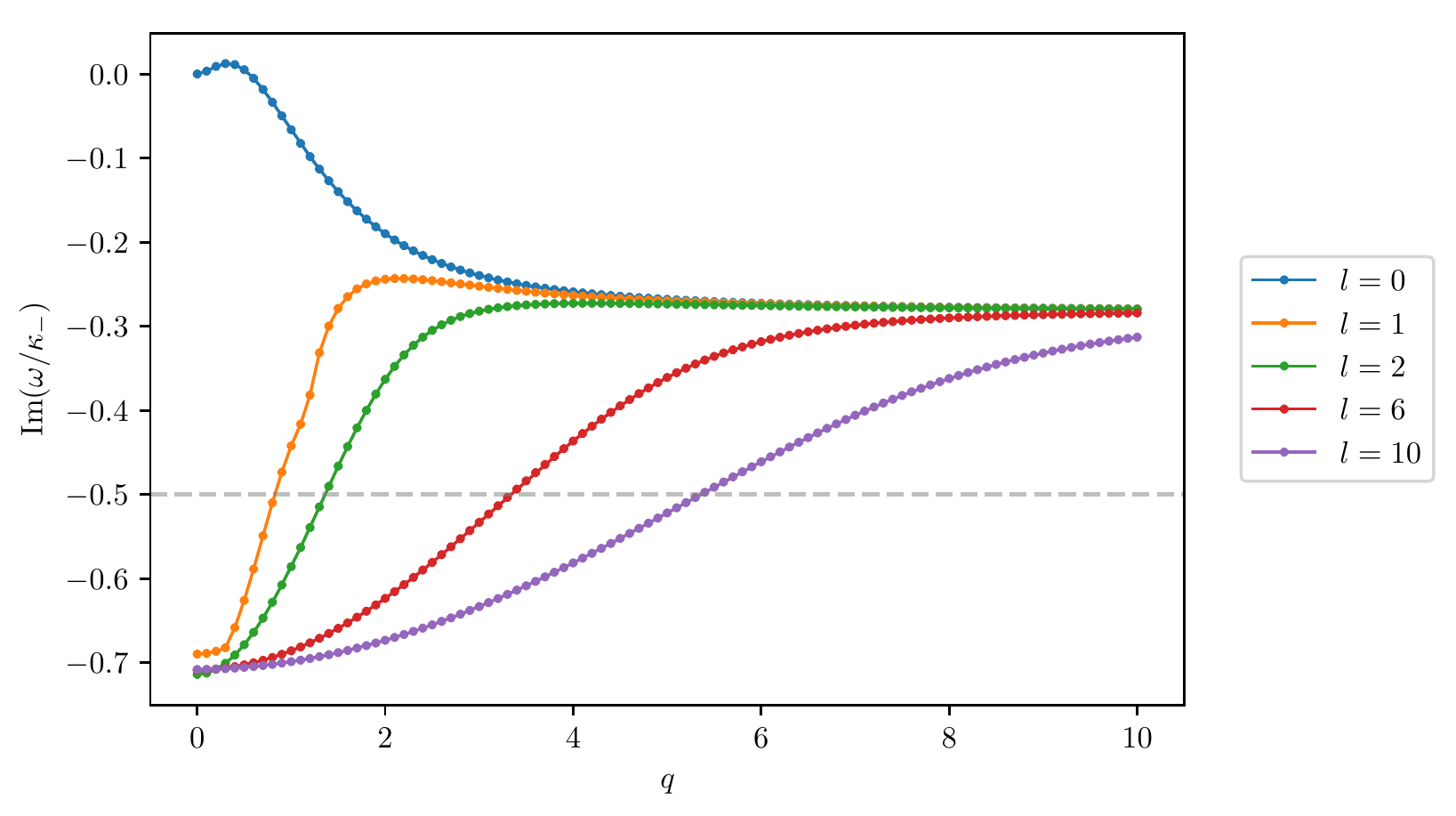}\\
  \caption{The lowest-lying QNMs for $\Lambda=0.02$ and $Q/Q_m=0.9950$.}\label{ds}
\end{figure}

We first present the lowest-lying QNMs in FIG.\ref{ds} for the case of $\Lambda=0.02$ and $Q/Q_m=0.9950$, where SCC would be violated by the $l=1$ dominant de-Sitter mode if the massless scalar field was uncharged. However, once we charge the scalar field, the $l=0$ trivial zero mode becomes non-trivial. In particular, it demonstrates the superradiant instability in the small charge regime and then stabilizes when the charge is large enough, which is consistent with the previous observation made in \cite{ZZPWA,KZ}. But no matter whether it is unstable or stable, this $l=0$ zero mode keeps lying well above the $-1/2$ horizontal threshold line, thus saves SCC out.

\begin{figure}
  \centering
  \includegraphics[width=12.0cm]{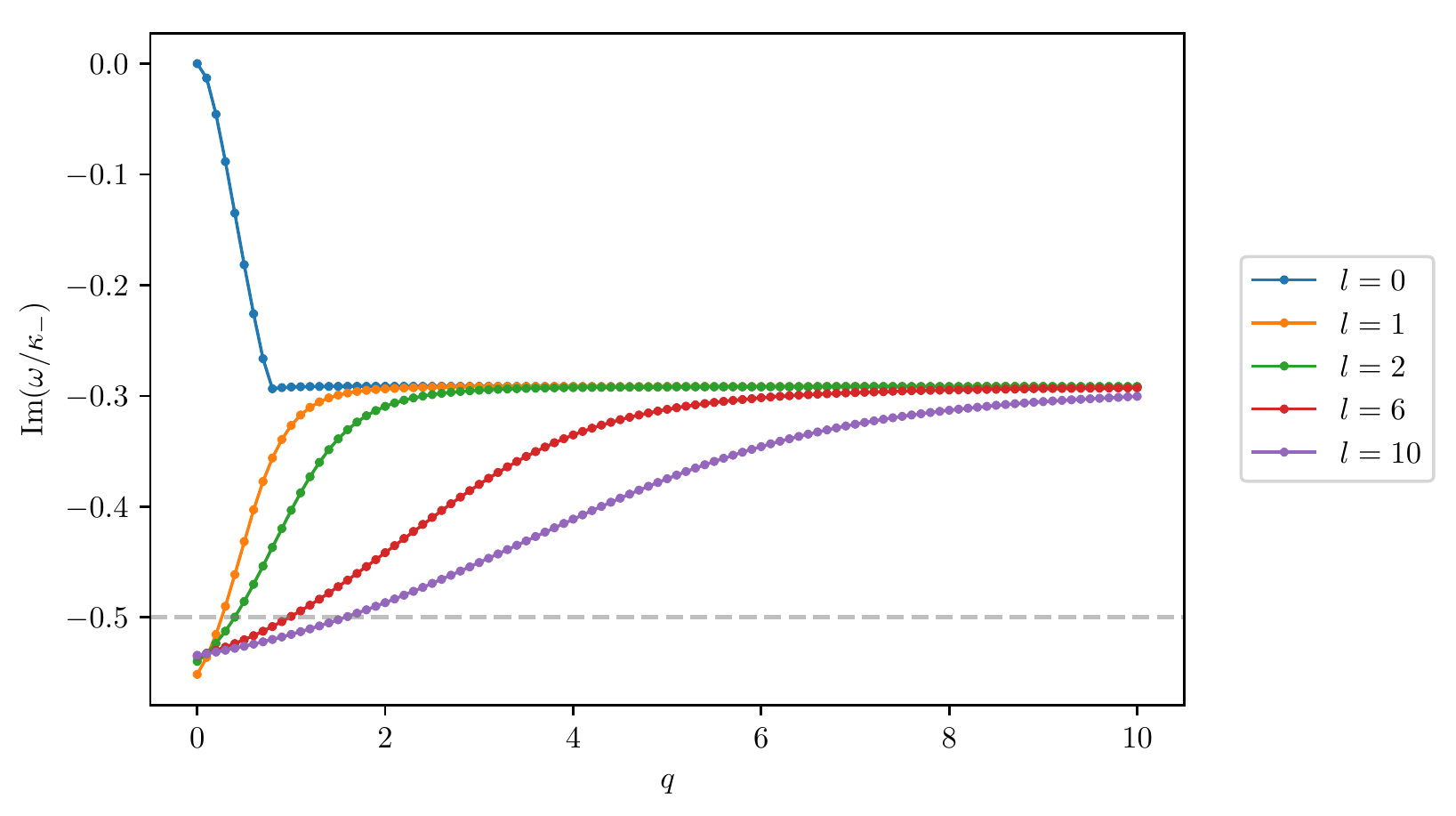}\\
  \caption{The lowest-lying QNMs for $\Lambda=0.14$ and $Q/Q_m=0.9950$.}\label{ps}
\end{figure}

\begin{figure}
  \centering
  \includegraphics[width=12.0cm]{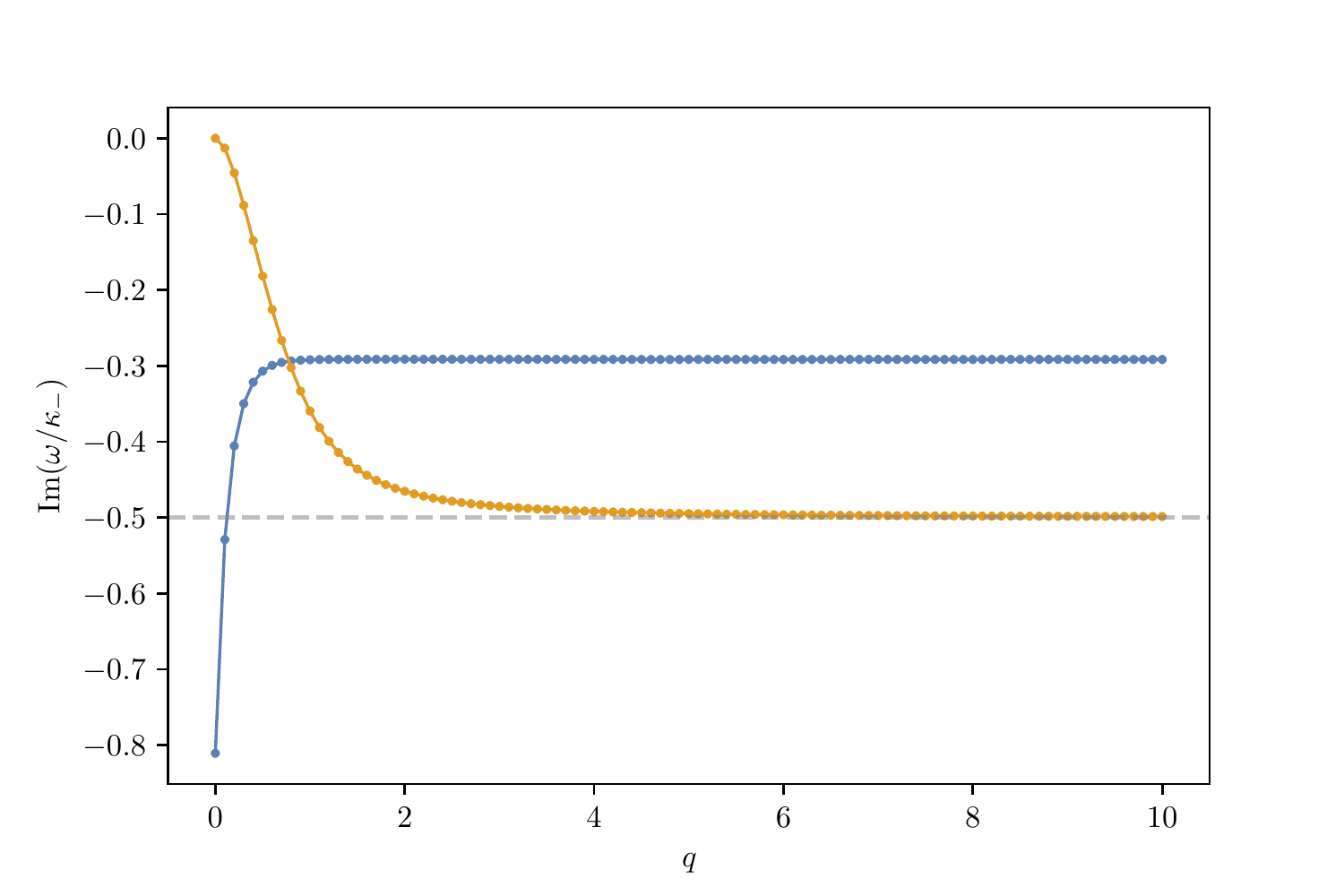}\\
  \caption{The $l=0$ dominant and sub-dominant QNMs for $\Lambda=0.14$ and $Q/Q_m=0.9950$, where the orange points and blue points denote the zero mode and near extremal mode, respectively.}\label{l0}
\end{figure}

Now let us turn to the case of $\Lambda=0.14$ and $Q/Q_m=0.9950$, where SCC would be violated by the photon sphere dominant mode if the massless scalar field was uncharged. As shown in FIG.\ref{ps}, the $l=0$ trivial zero mode still becomes non-trivial once the scalar field is charged, although it does not demonstrate the superradiance instability any more, consistent with the observation made in \cite{ZZPWA,KZ} that a large cosmological constant stabilizes the system. Similarly, at the presence of the charge, this mode keeps SCC from being violated again. As an aside remark, there is an obvious non-smoothness for the behavior of the $l=0$ dominant mode as one cranks up the charge. As demonstrated in FIG.\ref{l0}, such a non-smoothness arises from the fact that the $l=0$ near extremal mode takes over the dominant position from the $l=0$ zero mode when the charge is large enough.

\begin{figure}
  \centering
  \includegraphics[width=12.0cm]{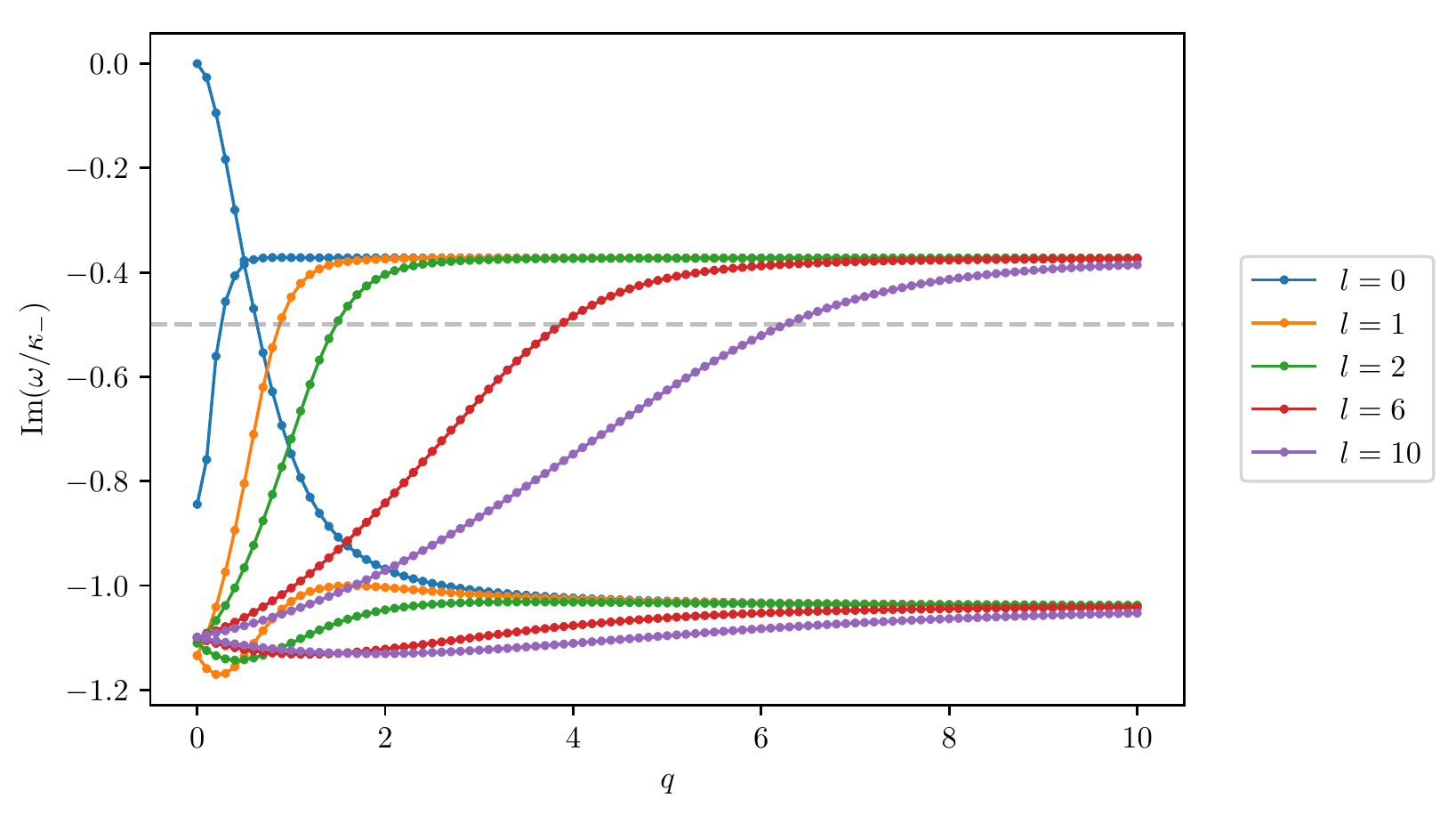}\\
  \caption{The dominant and sub-dominant QNMs for $\Lambda=0.14$ and $Q/Q_m=0.9985$.}\label{extremal}
\end{figure}

\begin{figure}
  \centering
  \includegraphics[width=12.0cm]{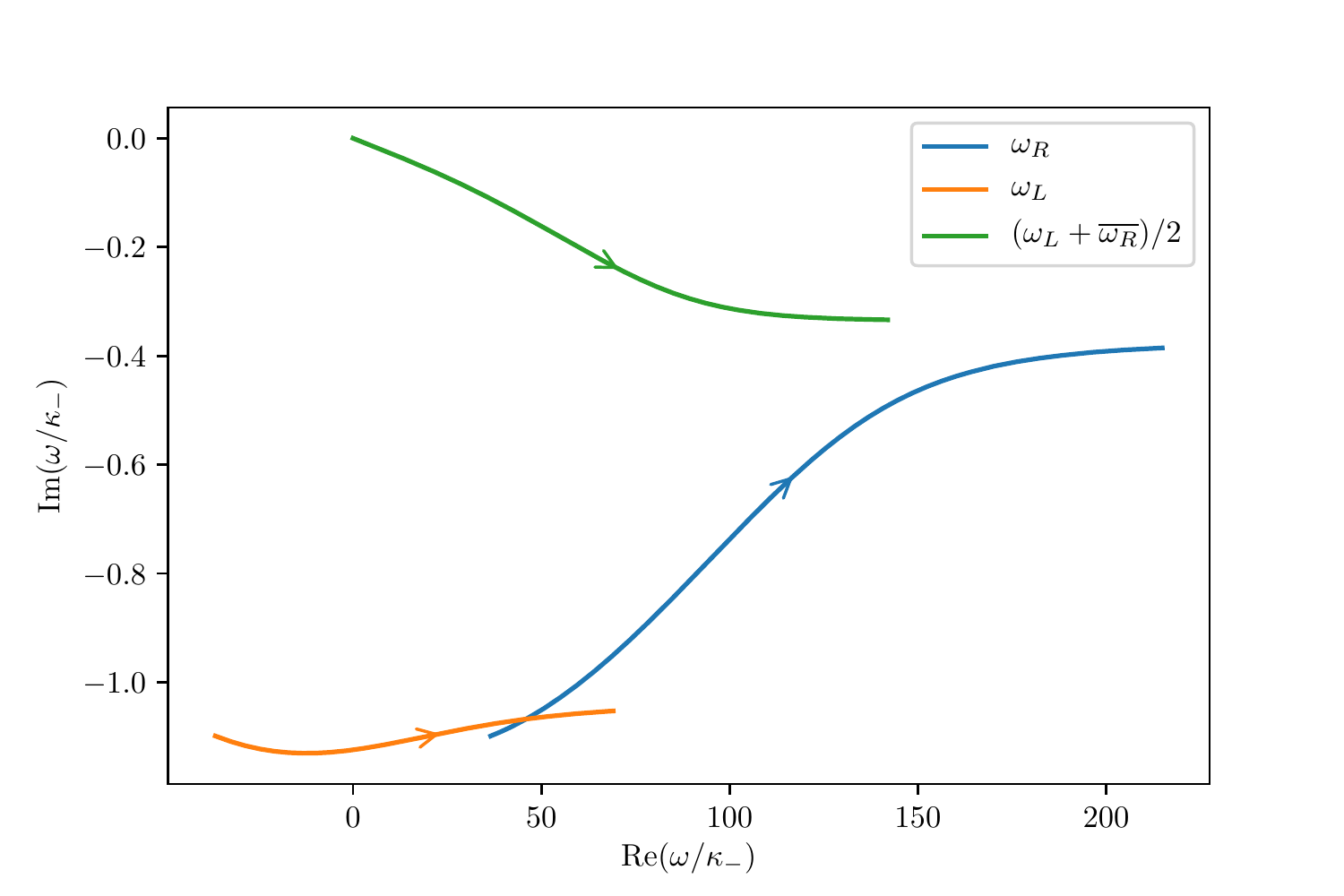}\\
  \caption{The symmetry breaking of the $l=10$ left and right photon sphere modes for $\Lambda=0.14$ and $Q/Q_m=0.9985$. The orange points denote the left photon sphere modes $\omega_L$, the blue points denote the right photon sphere modes $\omega_R$, and the green points denote $(\omega_L+\overline{\omega_R})/2$, where the arrow indicates the increase of the charge $q$ from $0$ to $10$.}\label{symmetry}
\end{figure}

Next let us see what happens to the case of $\Lambda=0.14$ and $Q/Q_m=0.9985$, where SCC would be violated by the $l=0$ dominant extremal mode if the massless scalar field was uncharged. As such, we would like to depict the corresponding dominant and sub-dominant QNMs in FIG.\ref{extremal}. As one can see, due to the $l=0$ non-trivial zero mode at the presence of the charge, there is no violation of SCC. In addition, it is noteworthy that the presence of the charge breaks the left and right symmetry between the $l\neq 0$ photon sphere modes with respect to the imaginary axis on the $\omega$-plane. To be more precise, as illustrated in FIG.\ref{symmetry} for the $l=10$ photon sphere modes, the magnitude of both of the real and imaginary parts of $\omega_L+\overline{\omega_R}$ increases from zero as the charge $q$ varies from $0$ to $10$.

\begin{figure}
  \centering
  \includegraphics[width=12.0cm]{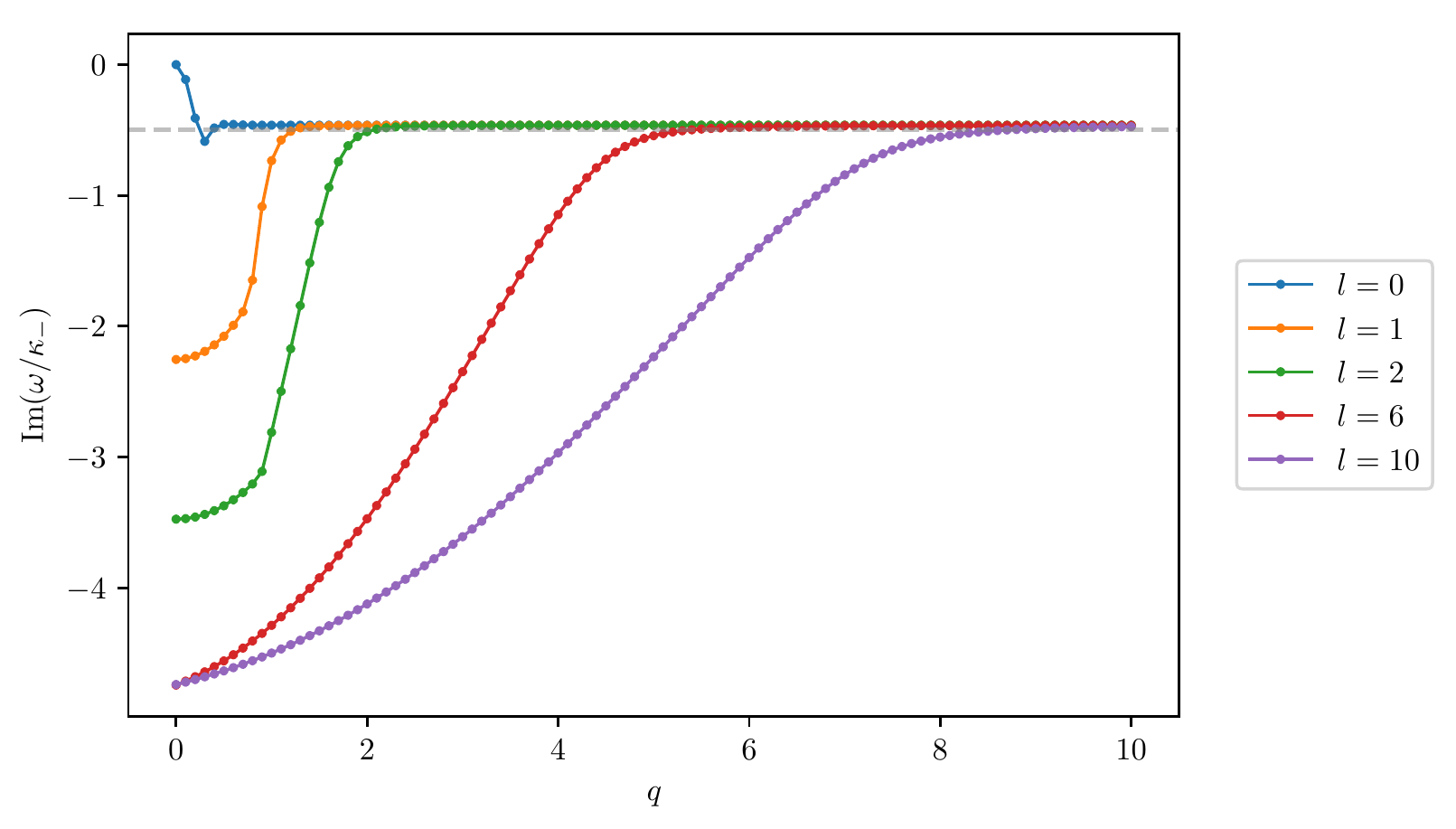}\\
  \caption{The lowest-lying QNMs for $\Lambda=0.14$ and $Q/Q_m=0.9999$.}\label{limit}
\end{figure}

\begin{figure}
  \centering
  \includegraphics[width=12.0cm]{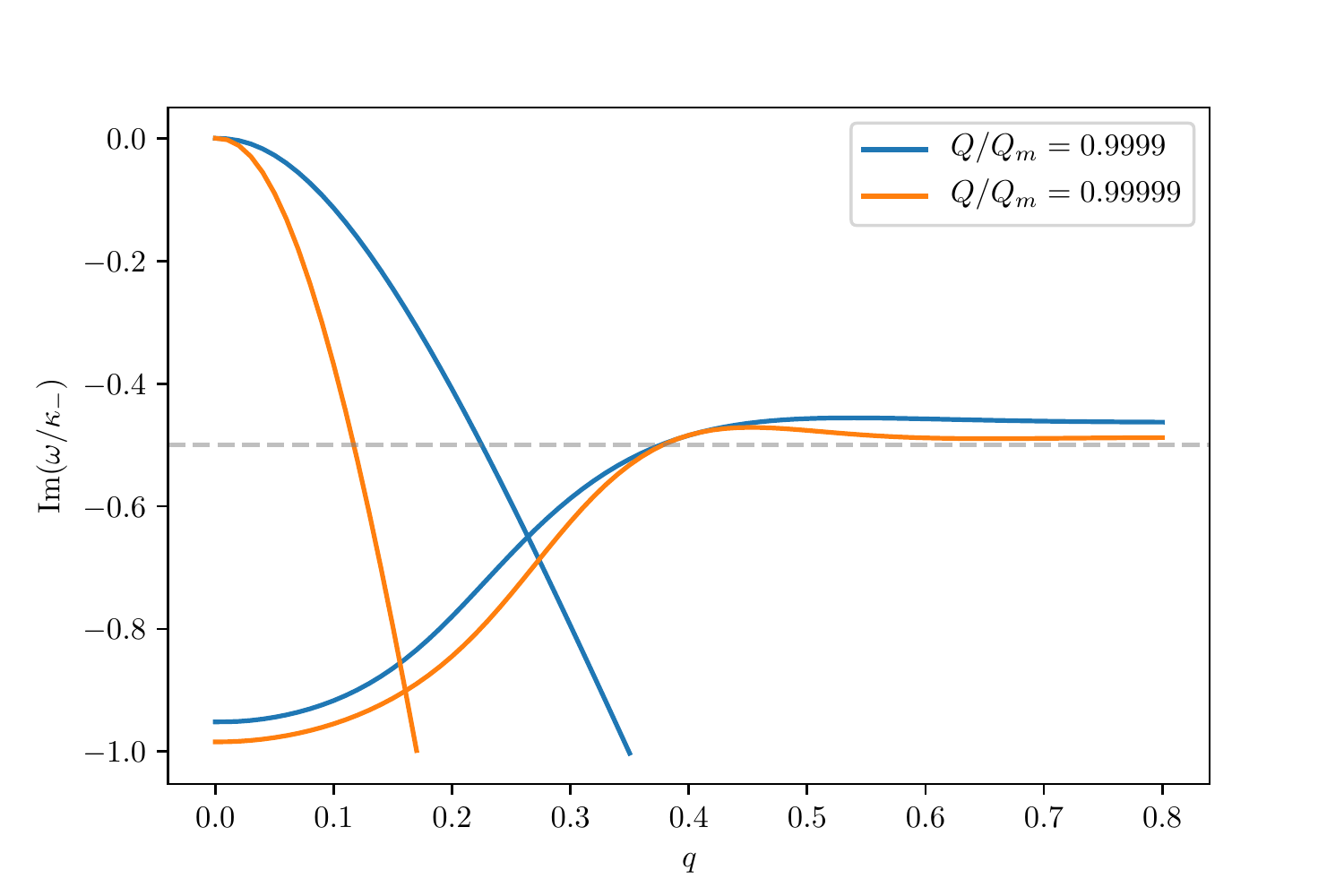}\\
  \caption{The violation regime for $\Lambda=0.14$ gets bigger when the extremal limit is approached.}\label{zoomin}
\end{figure}

Comparing FIG.\ref{ps} and FIG.\ref{extremal}, one can see that the minimum of $\frac{\text{Im}(\omega)}{\kappa_-}$ for the dominant QNMs gets decreased when the charge of black hole is increased toward the extremal limit. Actually this is the case. Moreover, SCC is still violated in such a limiting case\footnote{When finalizing our paper, we are informed by Aron Jansen that they do find there is still a violation of SCC when one goes to $Q\rightarrow Q_m$ limit in \cite{Aron}, and suggested to add such a limiting point in the moduli space for a complete picture. We are grateful to Aron Jansen and his companions for such an improvement.}. To see this explicitly, we like to depict the lowest-lying QNMs in FIG.\ref{limit} for $\Lambda M^2=0.14$ and $Q/Q_m=0.9999$. Obviously, although SCC is respected in the regime where the charge of the scalar field is sufficiently small or sufficiently large, the violation of SCC still occurs when the scalar field is appropriately charged within the albeit narrow regime. Furthermore, we zoom in this violation regime in FIG.\ref{zoomin}, which demonstrates this violation regime becomes bigger when one approaches the extremal limit. In particular, the minimal violation charge gets smaller while the maximal violation charge keeps almost unchanged. This indicates that the maximal violation charge may converge to a finite value in the extremal limit. Namely, once the charge of the scalar field is larger than this value, SCC is always respected\footnote{It is noteworthy that the wiggles found in \cite{DRS2} for some region of the moduli space may change such a picture.}.

\section{Conclusion and discussion}
To address SCC in the presence of the massless charged scalar field on top of the RNdS black hole, we have succeeded in calculating the corresponding low-lying QNMs by the time domain analysis. To this end, we generalize the characteristic formulation to the charged case in the double null coordinates. As a result, we find that the presence of the massless charged scalar field can recover SCC in the RNdS black hole except in the highly extremal limit, where the violation can still occur when the scalar field is appropriately charged.

In addition, among others, our numerical result also demonstrates two interesting patterns. First, The $l=0$ dominant mode seems to be always dominant over the $l\neq 0$ dominant modes. Second, it seems that the imaginary parts of different $l$ modes converge to some discrete values in the large $q$ limit. Both of these beg an analytic analysis. On the other hand, so far our investigation has been restricted at the linear level. So not only is it intriguing to see how the would-be Cauchy horizon becomes singular but also it is important to figure out the superradiant instability induced final state by the fully non-linear numerical simulation. We hope to address some of these topics in the near future.


\begin{acknowledgments}
This work is partially supported by NSFC with Grant No.11475179, No.11675015, and No.11775022, as well as by FWO-Vlaanderen through the project G020714N, G044016N, and G006918N. YT is also supported by the ``Strategic Priority Research Program of the Chinese Academy of Sciences" with Grant No.XDB23030000. BW acknowledges the support by NSFC with Grant No.11575109. HZ is supported by the Vrije Universiteit Brussel through the Strategic Research Program ``High-Energy Physics", and he is also an individual FWO Fellow supported by 12G3515N. We like to thank Vitor Cardoso, Kyriakos Destounis, and Aron Jansen for their numerous communications on various issues related to this project. Thanks are also due to Pau Figueras and Ran Li for their valuable discussions on our numerical scheme. We are also grateful to Shahar Hod, Jorge Santos, and Xiaoning Wu for their helpful discussions on SCC.
\end{acknowledgments}

\end{document}